# Tunable THz absorption in photonic crystal including graphene and metamaterial

M. Montaseri[1], M. Hosseini*[1], M. J. Karimi[1]

[1]Department of physic, Shiraz University of Technology, 313-71555 Shiraz, Iran

*Corresponding author (hosseini@sutech.ac.ir)

In this paper, a photonic crystal containing graphene and metamaterial layers is investigated. The absorption spectrum of the structure in the terahertz range is obtained using the transfer matrix method. The results show that by adding a Si, $SiO_2$ or metamaterial layer between two graphene layers, the terahertz absorption increases significantly. The results also reveal that in wide range of physical parameters, the approximately complete absorption (∼100%) occurs. Furthermore, the results indicate that the structure with metamaterial layer has the highest absorption performance.

## I. INTRODUCTION

In recent years, graphene-based materials have attracted much attention due to their physical characteristics and potential applications in electronic and optoelectronic devices, chemical sensors, nano-composites and energy storage [1]. The optical properties of the graphene can be tuned by adjusting of its plasmon frequency via manipulating the Fermi level using an externally applied electric field [2]. On the other hand, the light absorption plays an important role in design of photodetectors and photovoltaic devices. The perfect absorption has been achieved by different methods such as using i) sparseness and imperfect alignment of the vertical single-walled carbon nanotubes [3], ii) ordered periodic structures [4], iii) graphene-based structures [5].

So far, several researchers investigated the optical absorption of graphene-based structures. For example; Thongrattanasiri et al. have shown that a single sheet of doped graphene, patterned into a periodic array of nano-disks exhibits 100% light absorption [3] The tunability of absorption with the temperature in the graphene layer investigated by Ghasempour [6]. Zhu et al. have investigated the dielectric–graphene–metal groove-grating absorber and show that a high efficiency (95%) absorptive spectrum in near infrared range could be achieved by tuning the applied voltage on graphene [7]. Deng et al. have studied the THz absorption in graphene-based hetero structures and indicate that the THz absorption may be tuned continuously from 0 to 100 by adjusting the chemical potentials [8].

The tera hertz devices including absorbers, sensors and sources are subjects of many researches in the last years [9-13]. The graphene-based metamaterials are also good candidate for developing new absorber in terahertz (THz) [9] and infrared ranges [10]. Metamaterials are artificial materials that consist of sub-wavelength electric circuits and can exhibit exotic optical properties beyond those achievable in natural materials [11-13]. In this regard, Watts et al. have reported extensively the theory, characterization and implementation of metamaterial perfect absorbers [14]. Nefedov et al. have shown that 100% light absorption can be obtained in a graphene-based hyperbolic metamaterial [15]. Alaee et al. have introduced perfect absorbers based on graphene micro-ribbon metamaterial in far-infrared wavelength region [16]. Andryieuski et al. have designed a graphene based metamaterial absorber using the effective surface conductivity in the THz regime [17]. Linder et al. have studied the absorption properties of graphene-based anisotropic metamaterial structures and show that a strongly enhanced absorption appears overbroad range of incidence angles [18]. They also show that this is due to a coupling between light and wave propagating along the graphene-metamaterial structures.

In this paper, we propose a new structure to obtain tunable complete absorption in the terahertz region. The paper is organized as follows: We describe the theoretical framework in Section 2. Then, the results are discussed in Section 3, and finally, the conclusions are given in Section 4.

## II. THEORITICAL MODEL

As shown in Fig. 1, we consider a photonic crystal containing different layers: glasses with thickness $d_{gl}$ (gray regions), titanium dioxide with thickness $d_T$ (red), graphene with thickness $d_G$ (purple) and metamaterial, Si or SiO$_2$ with thickness $d_x$ (yellow region).

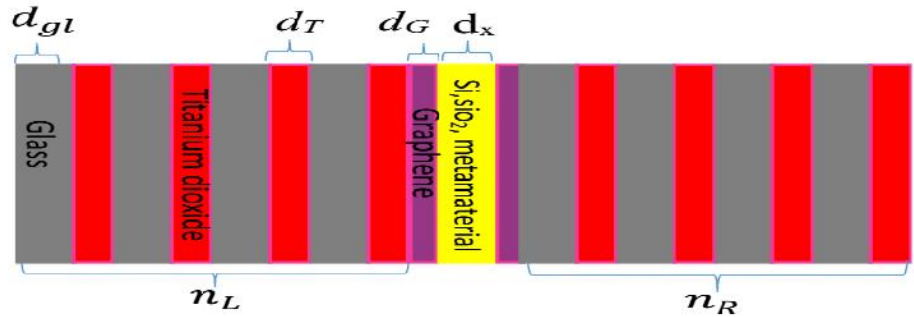

FIG. 1. A schematic diagram of our proposed photonic crystals.

Also, $n_L$ and $n_R$ are the number of the dielectric layers on the left and right sides of photonic crystal, respectively. The permittivity of graphene in the THz frequency range can be calculated from the Drude model [19, 20].

$$\varepsilon_g = 1 + \frac{i\sigma}{\varepsilon_0 \omega_0 d_g} \tag{1}$$

where $d_G$ is the thickness of the graphene layer, $\varepsilon_0$ is the permittivity of vacuum and    is the angular frequency. The conductivity    as a function of frequency is written as [3,21,22]:

$$\sigma = \frac{e^2}{\pi \hbar} \frac{K_B T}{\Gamma - i\omega}[\frac{\mu_c}{K_B T} + 2\ln(e^{-\frac{\mu_c}{K_B T}} + 1)] \tag{2}$$

where $K_B$ is the Boltzmann constant, $T$ is the temperature, $\Gamma$ is relaxation rate,    is the angular frequency and $\mu_c$ is the chemical potential. The transmission of the incident light beam from each layer is calculated using the transfer matrix of the layer, which is [6]

$$M = \begin{pmatrix} \cos(kd) & -i\frac{n}{\eta}\sin(kd) \\ -i\frac{n}{\eta}\sin(kd) & \cos(kd) \end{pmatrix} \tag{3}$$

where $\eta$ is the impedance of medium, $d$ is the thickness of each layer, $n$ is the refractive index and $k$ is the wave vector as follows,

$$K = (\omega/c)\sqrt{\varepsilon} \tag{4}$$

$\varepsilon$ is the permittivity of each layer, $\omega$ is the angular frequency and $c$ is the speed of light in vacuum. In this paper field intensity and the refractive index of each layer is constant.

The electric and magnetic fields in the adjacent layers related by the following equation:

$$\begin{pmatrix} E_i \\ H_i \end{pmatrix} = M_i \begin{pmatrix} E_{i+1} \\ H_{i+1} \end{pmatrix} \tag{5}$$

The transfer matrix of the hetero-structure consist of $n$-layer is given by:

$$M = M_1 M_2 ...., M_n = \begin{pmatrix} A & B \\ C & D \end{pmatrix} \quad (6)$$

where *A*, *B*, *C* and *D* are the elements of matrix *M*. The reflection coefficient *r( )* and the transmission coefficient *t( )* for the left-incident plane wave can be expressed as follows [6]

$$r(Š) = \frac{Ay_1 + By_1y_2 - C - Dy_2}{Ay_1 + By_1y_2 + C + Dy_2}$$

$$t(Š) = \frac{2y_1}{Ay_1 + By_1y_2 + C + Dy_2} \quad (7)$$

The reflectance and transmittance of the structure at frequency $\omega$ are defined as:

$$T(Š) = \frac{y_2}{y_1}|t(Š)|^2, \quad R(Š) = |r(Š)|^2 \quad (8)$$

where $\eta_1 = \sqrt{\frac{\eta_0}{n_1}}$ and $\eta_2 = \sqrt{\frac{\eta_0}{n_2}}$ are the impedance environment characters for the right and left of photonic crystals, respectively.

### III. RESULT AND DISCUSSION

The material parameters used in our calculations are as follows: the refractive indices of glass and titanium dioxide are 1.6 and 2.9, respectively which are approximately constant at THz frequencies [24]. Also, *T=300 ° K,*  *=2.5 meV/h, $\mu_c$=0.378 eV, $d_0$=70 μm, $d_g$=0.34 nm.* Except in results of Fig 2 and Fig. 7, $d_x$=1.0937*$10^{-5}$ m $n_L$

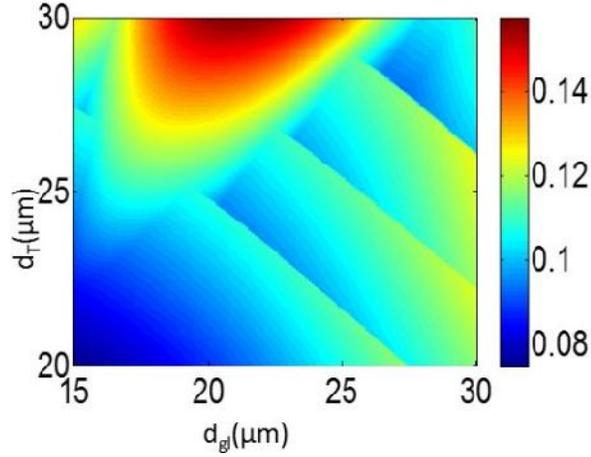

FIG. 2. The absorbance in case (1) for different thicknesses of the dielectric layers.

For more clarity, the results are presented in three cases. In case (1) there is nothing between graphene layers, in case (2) Si or SiO$_2$ placed between two graphene layers and $d_x$ layer filled with metamaterial in case (3).

Case (1): $d_x = 0. n_L = 4. n_R = 8$.

The absorbance of this case is plotted in Fig. 2 for different values of $d_T$ and $d_g$. This figure shows that the structure has a low absorbance and its maximum is about 15%.

Case (2): Si or SiO$_2$, $d_x = 20\,\mu$ . $n_L = 4. n_R = 8$.

Figs. 3(a) and 3(b) present the absorbance for the case when the $d_x$ layer filled with Si and SiO$_2$, respectively. This figure reveal that the absorbance increases with respect to the pervious case (Fig. 2) and that minimum is about 50% (60%) when the structure containing Si (SiO$_2$) layers. It is also seen that the higher absorbance region (dark-red region) of Fig. 3(b) is greater than that of Fig. 3(a).

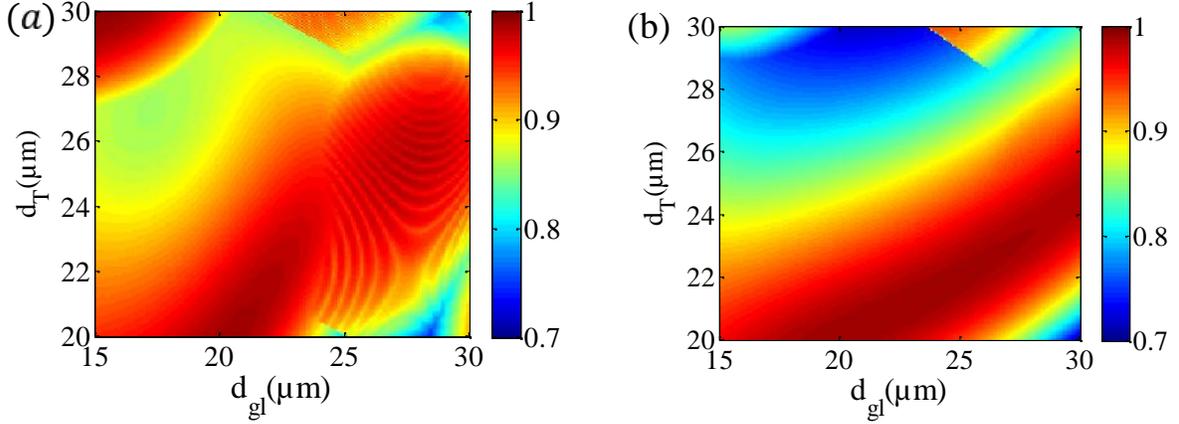

FIG. 3. The absorbance of the photonic crystal containing Si (a) and SiO2 (b). (Case 2)

Case (3): Metamaterial, $d_g = 20\,\mu$ , $d_T = 20\,\mu$

In Fig. 4, the absorbance of the photonic crystal versus the wavelength of incident light is plotted for different values of $n_L$ and $n_R$. In this figure the refractive index of metamaterial is -1.5. Fig. 4(a) shows that for a fixed value of $n_R$ (=8) the absorbance of $n_L$=5 is larger than that of other values of $n_L$. Also, the position of the absorption peak shifts to the higher values of the wavelength with increasing $n_L$. But, Fig. 4(b) reveals that for a fixed value of $n_L$ (=5), the absorbance for different values of $n_R$ has similar behavior and the same peak position.

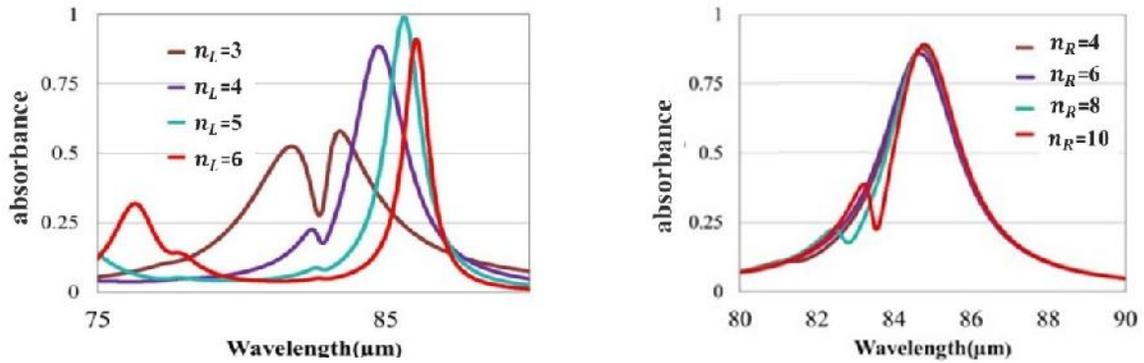

FIG. 4. Absorbance versus wavelength for different values of $n_L$ (a) and $n_R$ (b).

In Fig. 5. the absorbance and wavelength of the peak position are plotted versus $d_T$ and $d_g$ for $n_L = 5$ and $n_R = 8$. This figure shows that for most values of $d_T$ and $d_g$ the absorbance is larger than 90%. Also, the wavelength of the peak position is about in $60 - 120\ \mu$ range.

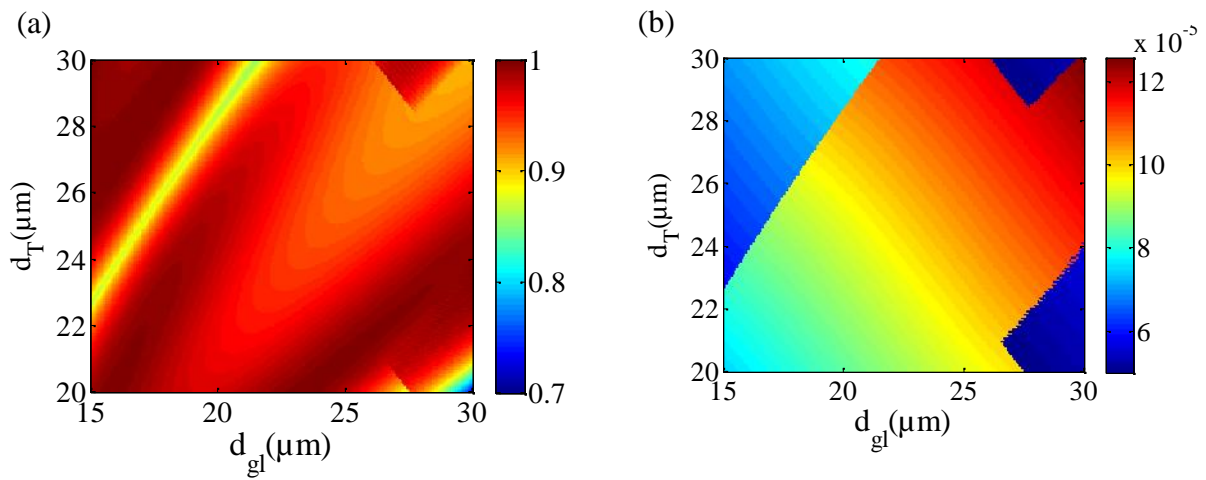

FIG. 5. Absorbance (a) and peak position wavelength (b) for different values of $d_T$ and $d_{gl}$ (Case 3).

The absorbance and peak position for different values of refractive index and metamaterial layer thickness are plotted in Fig. 6. From this figure, it is seen that the absorbance increases and reaches to 100% when the refractive index reaches to -1, especially for higher thickness of metamaterial layer. Fig 6 (b) shows that the wavelength of the peak position is about in range $72 - 76\,\mu$

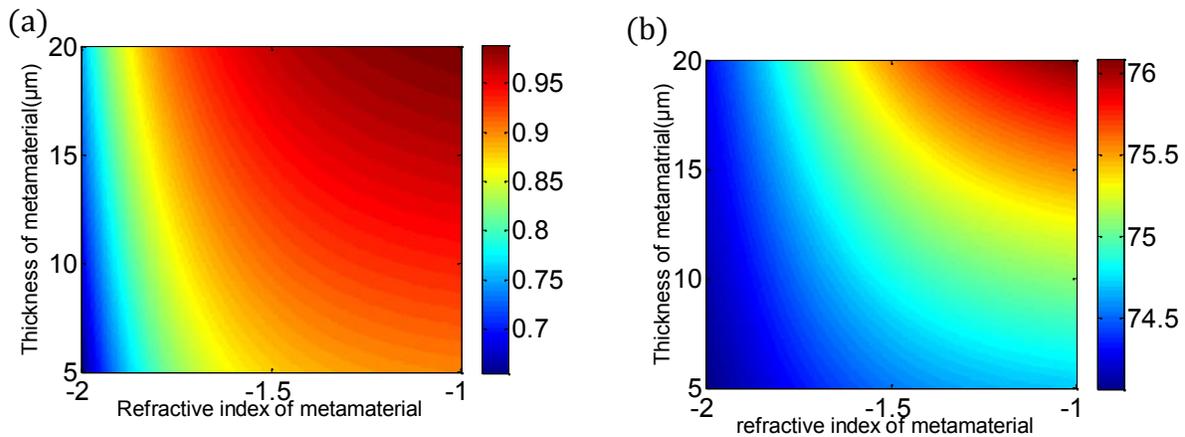

FIG. 6. Absorbance (a) and peak position wavelength (b) for different values of refractive index and thickness of metamaterial layer.

By comparing Figure 3 and 4 and Figure 5 and 6 can be seen that with presence in most parts of the metamaterial absorbs 100 percent but that is the structure of graphene without the metamaterial. Absorption is very low and this is different with The frequency of this structure with frequency when the metamaterial structure is present.

In Fig. the absorbance is shown for different values of $d$ and $d$ with $(n_L = 5. n_R = 8)$ .From this figure, it is seen that the absorbance of the case reaches to its maximum (100%). Thus, we can conclude that a perfect absorption can be obtained by suitable choice of layer thickness.

10## IV. CONCLUSION

In this work, we have proposed a graphene photonic crystal structure in three cases: i) no matter, ii) Si or SiO$_2$, iii) metamaterial between graphene layers. Then the effects of the thickness and number of layers, and wavelength of the incident radiation on the absorbance are investigated. Results show that absorbance increases by inserting the Si (SiO$_2$) layer in the structure. However, the presence of the metamaterial layer leads to the more increases of absorbance with respect to the other cases. Overall, our results indicate that a large absorbance in terahertz region can be achieved by suitable choosing the type of material, number and thickness of the layers.

## V. REFERENCES

[1] V. Singh, D. Joung, L. Zhai, S. Das, S. I. Khondaker, and S. Seal, **56**(8), 1178-1271 (2011).

[2] L. Ju, B. Geng, J. Horng, C. Girit, M. Martin, Z. Hao and F. Wang, Nat. nanotechnol, **6**(10), 630 (2011).

[3] M. I. Khan, M. S. Ahsan, and Z. H. Mahmood, 7th Int. Conf. on Informatics, Electronics & Vision (ICIEV), 245-249 (2018).

[4] P. B. Clapham, and M. C. Hutley, Nature, **244**(5414), 281 (1973).

[5] X. H. Deng, J. T. Liu, J. Yuan, T. B. Wang, and N. H. Liu, Opt. Express, **22**(24), 30177-30183 (2014).

[6] A. G. Ardakani, Eur. Phys. J B, **88**(7), 166 (2015).

[7] B. Zhu, G. Ren, S. Zheng, Z. Lin, and S. Jian, Opt. Commun. *308*, 204-210 (2013).

[8] X. H. Deng, J. T. Liu, J. Yuan, T. B. Wang, and N. H. Liu, Opt. Express, **22**(24), 30177-30183 (2014).